\documentclass[aps,prd,notitlepage,showpacs,nofootinbib,superscriptaddress]{revtex4-1}

\usepackage{graphicx}
\usepackage[utf8]{inputenc} 

\usepackage{amsmath}
\usepackage{amssymb}
\usepackage{float}
\usepackage{comment}
\usepackage{slashed}
\usepackage[normalem]{ulem}

\usepackage{caption}
\usepackage{subcaption}
\usepackage[usenames,dvipsnames]{color}
\usepackage[colorinlistoftodos]{todonotes}
\usepackage[colorlinks=true,citecolor=darkred,urlcolor=darkred, pdfborder={0 0 0}]{hyperref}
\usepackage[normalem]{ulem}

\definecolor{darkred}{rgb}{0.6,0,0}
\usepackage[colorinlistoftodos]{todonotes}

\definecolor{linkcolor}{rgb}{0,0,0.5}
 
\newcommand {\ignore}[1]{}

\def\gsim{\raise0.3ex\hbox{$\;>$\kern-0.75em\raise-1.1ex\hbox{$\sim\;$}}}
\def\lsim{\raise0.3ex\hbox{$\;<$\kern-0.75em\raise-1.1ex\hbox{$\sim\;$}}}

\newcommand{\sm}{{Standard Model }}

\providecommand{\be}{ \begin{equation} } 
\providecommand{\ee}{ \end{equation} }
\providecommand{\bea}{\begin{eqnarray}}
\providecommand{\eea}{\end{eqnarray}}
\providecommand{\nn}{\nonumber}

\providecommand{\to}{\rightarrow}

\usepackage{soul}

\definecolor{mightnightblue}{RGB}{25,25,112}

\definecolor{brown}{rgb}{0.59, 0.29, 0.0}

\def\21{$\mathrm{SU(2)_L \otimes U(1)_Y}$}

\def\sm{standard model }

\def\3311{$\mathrm{SU(3) \otimes SU(3)_L \otimes U(1)_X \otimes U(1)_{N}}$ }

\newcommand{\AddrAHEP}{%
  AHEP Group, Institut de F\'{i}sica Corpuscular --
  C.S.I.C./Universitat de Val\`{e}ncia, Parc Cient\'ific de Paterna.\\
 C/ Catedr\'atico Jos\'e Beltr\'an, 2 E-46980 Paterna (Valencia) - SPAIN}

\begin{document}


\title{\boldmath \color{BrickRed}  Scotogenic dark matter and Dirac neutrinos from unbroken gauged $B-L$ symmetry}

\author{ Julio Leite }\email{julio.leite@ific.uv.es}
\affiliation{\AddrAHEP}
\affiliation{Centro de Ci\^encias Naturais e Humanas, Universidade Federal do ABC,\\ 09210-580, Santo Andr\'e-SP, Brasil}

\author{América Morales}\email{america@fisica.ugto.mx }
\affiliation{Departamento de F\'isica, DCI, Campus Le\'on, Universidad de
Guanajuato, Loma del Bosque 103, Lomas del Campestre C.P. 37150, Le\'on, Guanajuato, M\'exico}

\author{Jos\'{e} W. F. Valle}\email{valle@ific.uv.es}
\affiliation{\AddrAHEP}

\author{Carlos A. Vaquera-Araujo}\email{vaquera@fisica.ugto.mx}
\affiliation{Consejo Nacional de Ciencia y Tecnolog\'ia, Av. Insurgentes Sur 1582. Colonia Cr\'edito Constructor, Del. Benito Ju\'arez, C.P. 03940, Ciudad de M\'exico, M\'exico}
\affiliation{Departamento de F\'isica, DCI, Campus Le\'on, Universidad de
Guanajuato, Loma del Bosque 103, Lomas del Campestre C.P. 37150, Le\'on, Guanajuato, M\'exico}

\begin{abstract}
\vspace{0.5cm}

We propose a simple extension of the \sm where neutrinos get naturally small ``scotogenic'' Dirac masses from an unbroken gauged $B-L$ symmetry, ensuring dark matter stability.  
The associated gauge boson gets mass through the Stueckelberg mechanism.
Two scenarios are identified, and the resulting phenomenology briefly sketched.

\end{abstract}

\maketitle
\noindent

\section{Introduction}

Amongst the major drawbacks of the Standard Model (SM) is the absence of neutrino mass and the lack of a viable dark matter candidate.
Amendments for both of these issues require new physics.
Particle dark matter candidates should be stable, at least on cosmological time scales~\cite{Bertone:2004pz}.
A simple way to ensure this is through the imposition of an adequate protecting symmetry whose nature is unknown.
For example, dark matter stability can result from a residual $\mathbb{Z}_{2}$ matter-parity symmetry~\cite{Alves:2016fqe}, 
from the R-parity symmetry in supersymmetric models~\cite{Barbier:2004ez} or from some $\mathbb{Z}_{n}$-like symmetry,
such as quarticity~\cite{Chulia:2016ngi,Chulia:2016giq,CentellesChulia:2017koy}. 
Although these in general are \emph{ad hoc} assumptions, it could be that they can follow naturally from the spontaneous breaking of an extended gauge 
symmetry~\cite{Dong:2017zxo,Kang:2019sab,Leite:2019grf}.

A specially attractive possibility is that neutrino mass and dark matter have a common origin, i.e. the same physics being responsible for both.
For example, dark matter could be mediator of neutrino mass generation~\cite{Ma:2006km, Farzan:2012sa, Hirsch:2013ola,Merle:2016scw}. 
Also the symmetry stabilising dark matter could be closely related to neutrinos.
For example, it could be an unbroken subgroup of the flavour symmetry that helps understand the neutrino oscillation parameters~\cite{Hirsch:2010ru, Boucenna:2011tj, Morisi:2012fg,Bonilla:2017ekt}.
In some cases this will lead to Dirac neutrinos, obtained as a consequence of flavour symmetry imposition~\cite{Aranda:2013gga,CentellesChulia:2018gwr}.

Likewise, the dark matter stabilising symmetry could be a residual $\mathbb{Z}_{n}$ subgroup of lepton number symmetry or $B-L$.
This may, again, lead to Dirac neutrinos. Such a possibility was explored in \cite{Bonilla:2018ynb}, assuming that $B-L$ is spontaneously broken down to a $\mathbb{Z}_n$ symmetry stabilising dark matter, as well as in \cite{Ma:2019coj}, where the residual dark matter stabilising symmetry follows from the soft breaking of $B-L$.

On the other hand, conservation of the full ungauged $B-L$ symmetry could stabilise dark matter.
A scenario of this type has been suggested within a bound-state dark matter scenario~\cite{Reig:2018mdk}. 
Indeed, this is a justified hypothesis since, despite decade-long searches~\cite{Agostini:2019hzm}, there has been no experimental evidence of $B-L$ breakdown.

In this letter we consider the alternative case of gauged unbroken $B-L$ as the dark matter stabilisation symmetry. 
The promotion of the accidental $B-L$ global symmetry of the \sm to a local one stands out for its simplicity, 
since the inclusion of three right-handed neutrinos $\nu_{iR}$ is enough to make it anomaly free and hence consistent.
Clearly $B-L$ preserving models are viable provided the associated $Z'$ boson develops an adequate mass. 
Here we study a Stueckelberg \cite{Ruegg:2003ps} $B-L$ extension of the \sm with naturally small neutrino masses.
These are achieved through the scotogenic approach, while the unbroken $B-L$ symmetry is
responsible for both the Dirac nature of the neutrino mass and the stabilisation of a dark matter candidate.

The letter is organized a follow. In Sec.~\ref{sec:model} we describe the theoretical setup, in Sec.~\ref{sec:scalar-spectrum} we discuss the scalar sector
and in Sec.~\ref{sec:stueck-mech} we describe the Stueckelberg mechanism, while in Sec.~\ref{sec:scot-neutr-mass} we give the mass generation mechanisms
in the two alternative realisations of our scenario.
Finally, in Sec.~\ref{sec:conclusions} we briefly comment on the phenomenology and summarize. 

\section{Two scenarios}
\label{sec:model}

We start from the basic setup provided by the \sm fermion and scalar sectors, as defined in Table \ref{tabSM}.
This has an automatic global ``baryon number minus lepton number'' symmetry, we call simply $U(1)_{B-L}$.
This symmetry, however, cannot be directly promoted to a local one, as it exhibits non-vanishing $[U(1)_{B-L}]^3$ and $[Grav]^2\times[U(1)_{B-L}]$ anomalies.
Therefore, in order to ``gauge'' $U(1)_{B-L}$ consistently, we need to extend the \sm field content so as to ensure anomaly cancellation
~\footnote{Since the pioneer paper of Pati and Salam~\cite{Pati:1974yy} there have been many suggestions for gauging $B-L$ just as a U(1) symmetry see, e.g.,
  Refs.~\cite{Davidson:1978pm,Marshak:1979fm,Malinsky:2005bi} also within a Stueckelberg approach~\cite{Feldman:2011ms}.}.
\begin{table}[h]
\centering
\begin{tabular}{|c|c|c|c|c|}
\hline
Fields & $SU(3)_C$ & $SU(2)_L$ & $U(1)_Y$ & $U(1)_{B-L}$ \\
\hline\hline
$L_{iL}$ & 1 & 2 & --1/2 & --1  \\
$e_{iR}$ & 1 & 1 & --1 & --1 \\
$Q_{iL}$ & 3 & 2 & 1/6 & 1/3  \\
$u_{iR}$ & 3 & 1 & 2/3 & 1/3 \\
$d_{iR}$ & 3 & 1 & --1/3 & 1/3 \\
\hline
\hline
$H$ & 1 & 2 & 1/2 & 0  \\
\hline
\end{tabular}
  \caption{Standard Model fermions and scalars and their gauge transformation properties and the global $B-L$.}
    \label{tabSM}
\end{table}

  A simple way to achieve this is by adding three right-handed fermions, $\nu_{iR}$, singlets under the SM gauge group but with a $B-L$ charge of $-1$.
  Adding the $\nu_{iR}$ allows for a tree-level Dirac mass for neutrinos through the term  $y\overline{L_L}\tilde{H} \nu_{iR}$.
  Viable neutrino masses require very tiny Yukawa couplings, $y \lesssim \mathcal{O}(10^{-11})$.
  On the other hand, if the $\nu_{iR}$ have a Majorana mass term, the $B-L$ symmetry is broken, and the active neutrinos get masses via the type-I seesaw mechanism,
  requiring the $B-L$ scale to be around the unification scale for Yukawa couplings of order 1.

  Here we propose alternative ways of generating small neutrino masses, relying neither on unnaturally suppressed Yukawa couplings nor on inaccessibly large energy scales.
 At the same time, in our scenario dark matter candidates mediate neutrino mass generation in a scotogenic fashion, and are stabilised by the conserved $B-L$ symmetry.
 We suggest two \sm extensions with gauged $B-L$ and dark matter as the mediator of neutrino mass generation.  
  
  In Model A, we add a new doublet $\eta$ and a singlet $\sigma$ in the scalar sector, both charged under the $B-L$ symmetry.
  In the fermion sector, three right-handed neutrinos, $\nu_{iR}$, as well as three gauge singlet fermions, $S_{iR}$, are introduced.
  The particle content and symmetry properties are given in Table~\ref{tabA}.
  In addition to the gauge symmetries, a $\mathbb{Z}_2$ is imposed, under which all the standard fields in Table \ref{tabSM} transform trivially.
  Notice that, even though there is a Majorana mass term for the three 2-component SM singlet fermions $S_R$
  \footnote{Notice that we stick to the chirally-projected 4-component description for the intrisically 2-component electrically neutral fermions~\cite{Schechter:1980gr}.}, it conserves $B-L$.
As a result, this is consistent with the Dirac nature of the light neutrinos.
 
We also propose a variant, Model B, shown in Table {\ref{tabB}}. This differs from Model A due to introduction of three extra fermion fields: $S_L$.
In contrast to Model A, the charges of the new fields, except for those of $\nu_{iR}$, are not fixed but defined by a single integer $n\neq 0$.
Notice that, although $S_L$ and $S_R$ are SM singlets, they carry nonzero $B-L$ charges. 
\begin{table}[h]
\centering
\begin{tabular}{|c|c|c|c|c|c|}
\hline
Fields & $SU(3)_C$ & $SU(2)_L$ & $U(1)_Y$ & $U(1)_{B-L}$ & $\mathbb{Z}_{2}$ \\
\hline\hline
$\nu_{iR}$ & 1 & 1 & 0 & --1 & --  \\
$S_{iR}$ & 1 & 1 & 0 & 0  & + \\
\hline
\hline
$\eta$ & 1 & 2 & 1/2 & 1 & +  \\
$\sigma$ & 1 & 1 & 0 & 1  & -- \\
\hline
\end{tabular}
  \caption{SM extension A: new fields and their symmetry properties.}
    \label{tabA}
\end{table}
\begin{table}[h]
\centering
\begin{tabular}{|c|c|c|c|c|c|}
\hline
Fields & $SU(3)_C$ & $SU(2)_L$ & $U(1)_Y$ & $U(1)_{B-L}$ &
$\mathbb{Z}_{2}$ \\
\hline\hline
$\nu_{iR}$ & 1 & 1 & 0 & --1 & --  \\
$S_{iL}$ & 1 & 1 & 0 & 2$n$  & + \\
$S_{iR}$ & 1 & 1 & 0 & 2$n$  & + \\
\hline
\hline
$\eta$ & 1 & 2 & 1/2 & 2$n$+1  & +  \\
$\sigma$ & 1 & 1 & 0 & 2$n$+1  & -- \\
\hline
\end{tabular}
  \caption{SM extension B: new fields and their symmetry properties. Here, $n( \neq0) \in \mathbb{Z}$.}
    \label{tabB}
\end{table}

In both cases, the $\mathbb{Z}_{2}$ symmetry in Tables \ref{tabA} and \ref{tabB} prevents the  appearance of $\overline{L_{iL}}\tilde{H}\nu_{iR}$ in the Yukawa sector.
As a result there are no tree-level neutrino masses.
Neverteless, this symmetry is broken in the scalar sector, making it possible for neutrinos to get calculable Dirac masses at the one-loop level. 

Notice that the $U(1)_{B-L}$ remains exactly conserved. 
%
%
This implies that the $\mathbb{Z}_2$ group, called matter parity, generated by $M_P = (-1)^{3(B-L)+2s}$, where $s$ is the field's spin,
is also exactly conserved.
Under $M_P$, the SM fields and $\nu_{iR}$ transform trivially. On the other hand, $S$, $\eta$ and $\sigma$, in Tables \ref{tabA} and \ref{tabB}, are $M_P$-odd fields.
Therefore, the lightest among them is stable by matter parity and can play the role of dark matter.
In either Model A or B, the dark matter candidate can be scalar or fermionic. When fermionic, dark matter would be Majorana-type in Model A and Dirac-type in Model B.

\section{Scalar spectrum}
\label{sec:scalar-spectrum}

Taking into account the fields and symmetries in Tables~\ref{tabA} and \ref{tabB}, the simplest scalar potential, shared by the two schemes sketched above, can be written as
\bea\label{V}
V &=& \sum_{s=H,\eta,\sigma} \left[ \mu_s^2 (s^\dagger s) + \lambda_s (s^\dagger s)^2\right] +  \lambda_{H\eta}(H^\dagger H)(\eta^\dagger \eta)+  \lambda^\prime_{H\eta}(H^\dagger \eta)(\eta^\dagger H)\\
&&+ \lambda_{H\sigma}(H^\dagger H)(\sigma^* \sigma)+  \lambda_{\eta\sigma}(\eta^\dagger \eta)(\sigma^* \sigma) + \frac{\mu_3}{\sqrt{2}}( \eta^\dagger H \sigma + h.c.)\,, \nn
\eea
where the last term breaks the $\mathbb{Z}_{2}$ symmetry softly, with the mass parameter $\mu_3$ assumed to be real for simplicity. 

Assuming that $B-L$ is not broken, it is easy to see that the matter parity of the new scalars is the opposite of that of the standard Higgs doublet $H$.
As a result, $M_P$ conservation requires the new scalars $\eta$ and $\sigma$ not to acquire any vev, and hence they do not mix with $H$. 
Therefore, similar to the \sm case, when the neutral component of $H$ acquires a vev, the $SU(2)_L\otimes U(1)_Y$ group is broken down to a $U(1)_Q$ subgroup, generated by the conventional electric charge operator $Q=T_3 + Y$. 
The CP-even field in the neutral component of $H$ becomes massive with $m_{h^0}^2 = 2\lambda_H v^2$ and is identified with the $125$ GeV Higgs boson observed at the LHC in 2012. 
The remaining components of $H$ are absorbed by the gauge sector, through the Higgs mechanism, making the $W^\pm$ and $Z$ vector bosons massive. 

The other scalars are in the $M_P$-odd or ``dark sector'' and do not acquire a vev. The first component of the scalar doublet $\eta$ corresponds to a massive charged scalar field, $\eta^\pm$, whose mass is
\be \label{ChScM}
m_{\eta^\pm}^2 = \frac{\lambda_{H\eta} v^2}{2} + \mu_{\eta}^2 .
\ee
The spontaneous symmetry breaking through $\langle H \rangle$ induces a mixing between the second component of $\eta$, $\eta^0$, 
and the singlet $\sigma$, arising from $(\sigma,\,\eta^0 )\, M^2_{\varphi}\,(\sigma,\,\eta^0 )^\dagger$, where
\be \label{CNScMM}
M^2_{\varphi}= \frac{1}{2}\left(
\begin{array}{cc}
 2\mu_\sigma^2 + \lambda_{H\sigma} v^2  & \mu_{3} v \\ \mu_{3} v  & 2\mu_\eta^2 + (\lambda_{H\eta}+\lambda^\prime_{H\eta})  v^2 \\
\end{array}
\right)\,.
\ee
Upon diagonalising the mass matrix above, we find two complex neutral scalars in the spectrum
\bea \label{CNSc}
\begin{pmatrix} \varphi_{1}^0 \\ \varphi_{2}^0 \end{pmatrix} = \begin{pmatrix} \cos \theta & \sin \theta\\ -\sin \theta & \cos \theta \end{pmatrix} \begin{pmatrix} \sigma\\ \eta^0 \end{pmatrix},\,\,\,\,\mbox{with}\,\,\,\, 2\theta = \arctan (\epsilon)=\arctan \left[\frac{2 \mu_3 v}{2(\mu_\sigma^2-\mu_\eta^2) + (\lambda_{H\sigma}-\lambda_{H\eta}-\lambda^\prime_{H\eta})v^2 }\right]\,.
\eea 
Notice that when the $\mathbb{Z}_{2}$-soft-breaking term, $\mu_3$, goes to zero, the mixing angle $\theta$ also vanishes. The mass eigenvalues associated with such states are
\bea \label{CNScM}
m_{\varphi^0_{(1,2)}}^2 &=& \frac{1}{4}\bigg\{2 (\mu_\eta^2 + \mu_\sigma^2)+ v^2 (\lambda_{H\eta} + \lambda^\prime_{H\eta} + \lambda_{H\sigma})  \mp \mathcal{F} \sqrt{\left[2(\mu_\sigma^2-\mu_\eta^2 ) + v^2( \lambda_{H\sigma}-\lambda_{H\eta} - \lambda^\prime_{H\eta})\right]^2+4\mu_3 v^2} \bigg\}\,,
\eea
respectively, where $\mathcal{F}=1$ for $(M_\varphi^2)_{22}/(M_\varphi^2)_{11}>1$ and $\mathcal{F}=-1$ otherwise.

It is worth noticing that in Eq.~(\ref{CNScMM}), the real and imaginary parts of the scalar fields appear together, that is, real and imaginary parts are degenerate in mass.
That means that, if dark matter is scalar, it is described by a complex field, in contrast to conventional scotogenic 
scenarios~\cite{Ma:2006km,Merle:2016scw} in which they are nearly degenerate but not exactly so.

\section{Stueckelberg Mechanism}
\label{sec:stueck-mech}

Neglecting kinetic mixing~\footnote{Kinetic mixing has been discussed in Refs.~\cite{Holdom:1985ag,Feldman:2007wj,Williams:2011qb}.}, after electroweak spontaneous symmetry breaking the $Z'$ boson associated with $U(1)_{B-L}$ remains unmixed with the standard EW gauge bosons. Its massive nature can be described by the kinetic Lagrangian \cite{Ruegg:2003ps}
\begin{equation}
\mathcal{L}^{\mathrm{St}}_{\mathrm{kin}}=-\frac{1}{4}Z'^{\mu\nu}Z'_{\mu\nu}+\frac{1}{2}(M_{Z'}Z'^{\mu}-\partial^\mu A)^2,
\end{equation}
which is invariant under the $U(1)_{B-L}$ gauge transformations
\begin{equation}
\begin{split}
Z'^{\mu}& \to Z'^{\mu}+\partial^\mu\Lambda,\\
A& \to A+ M_{Z'}\Lambda,
\end{split}
\end{equation}
where $A$ is a scalar Stueckelberg compensator and $Z'^{\mu\nu}=\partial^\mu Z'^\nu-\partial^\nu Z'^{\mu}$. 
Upon gauge-fixing, implemented by the $R_\xi$ gauge term
\begin{equation}
\mathcal{L}^{\mathrm{St}}_{\mathrm{fg}}=-\frac{1}{2\xi}(\partial_{\mu} Z'^{\mu}+M_{Z'} \xi A)^2,
\end{equation}
the $Z'$ boson acquires mass $M_{Z'}$  and the auxiliary field $A$ decouples, as
\begin{equation}
\mathcal{L}^{\mathrm{St}}_{\mathrm{kin}}+\mathcal{L}^{\mathrm{St}}_{\mathrm{fg}}=-\frac{1}{4}Z'^{\mu\nu}Z'_{\mu\nu}+\frac{1}{2}M^2_{Z'}Z'^{\mu}Z'_{\mu}-\frac{1}{2\xi}(\partial_{\mu} Z'^{\mu})^2+\frac{1}{2}\partial^\mu A \partial_\mu A -\frac{1}{2}M^2_{Z'} \xi A^2,
\end{equation}
up to a total derivative. Here $M_{Z'}$ is a free parameter of the model, unrelated to any vev and disconnected from the neutrino mass generation mechanism. 

The relevant $Z'$ interactions  are
\begin{eqnarray}\label{ZpInt}
\mathcal{L}^{Z'}_{\mathrm{f}}&=& g' Z'_\mu\sum_{i=1}^{3}\left[\frac{1}{3}(\overline{u}_i\gamma^\mu u_i+\overline{d}_i\gamma^\mu d_i)-\overline{e}_i\gamma^\mu e_i-\overline{\nu}_i\gamma^\mu \nu_i+2n\overline{S}\gamma^\mu S\right],\\
\mathcal{L}^{Z'}_{\mathrm{s}}&=&i g'(2n+1) Z'_\mu\left[ \eta^{-}\partial^\mu \eta^{+}-\eta^{+} \partial^\mu\eta^{-} +\sum_{i=1}^{2}\left(\varphi^{0*}_i\partial^\mu \varphi^{0}_i- \varphi^{0}_i\partial^\mu\varphi^{0*}_i \right)\right] \nn\\&&+ g'^{2}(2n+1)^2 {Z'}^{\mu}{Z'}_{\mu}\left( \eta^{-}\eta^{+} + \sum_{i=1}^{2}\varphi^{0*}_i\varphi^{0}_i\right),\nn\\
\mathcal{L}^{Z'}_{\mathrm{s+g}}&=& 2 e g' (2n+1) Z'^{\mu }\left\{ \left[A_\mu+ \cot (2\theta_W) Z_\mu\right]\eta^{-}\eta^{+} -\csc (2\theta_W) Z_\mu \left|\varphi^{0}_1 \cos \theta -\varphi^{0}_2 \sin \theta \right|^2\right.\nn\\
&&\left.+ \frac{\csc \theta_W}{\sqrt{2}} \left[W^+_\mu\eta^-(\varphi^{0}_1 \cos \theta -\varphi^{0}_2 \sin \theta)+\mathrm{ h.c.}\right]\right\},\nn
\end{eqnarray}
with $n=0$ in Model A and $n(\neq 0) \in \mathbb{Z}$ in Model B.

There are no gauge-mediated flavour-changing neutral currents and both $B-L$ and $M_P$ remain unbroken to all orders in perturbation theory, preserving the Dirac nature of neutrinos.

\section{Scotogenic neutrino masses} 
\label{sec:scot-neutr-mass}

We now give the most general renormalisable Yukawa Lagrangians for our models.
According to Tables \ref{tabA} and \ref{tabB} they can be written, respectively, as follows
\bea\label{Yuk}
-\mathcal{L}_Y^A &=& y^e \overline{L_L} H e_R + y^\nu \overline{L_L} \tilde{\eta} S_R + h \overline{(S_R)^c} \sigma \nu_R + \frac{1}{2}M_S^M\overline{(S_R)^c}S_R +  h.c. \,,\\
-\mathcal{L}_Y^B &=& y^e \overline{L_L} H e_R + y^\nu \overline{L_L} \tilde{\eta} S_R + h \overline{S_L} \sigma \nu_R + M_S^D\overline{S_L}S_R +  h.c. \,,\nn
\eea
where the flavour indices have been omitted.

\begin{figure}[h]
\centering
\includegraphics[scale=.8]{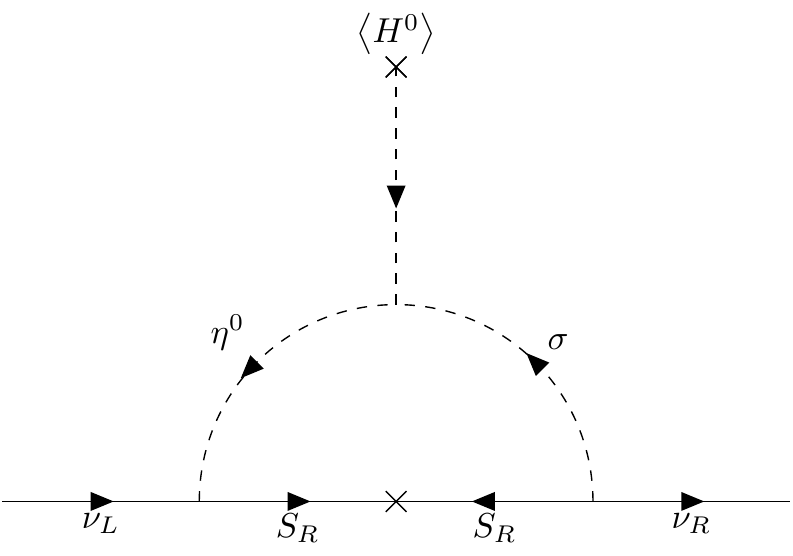}~~~~~
\includegraphics[scale=.8]{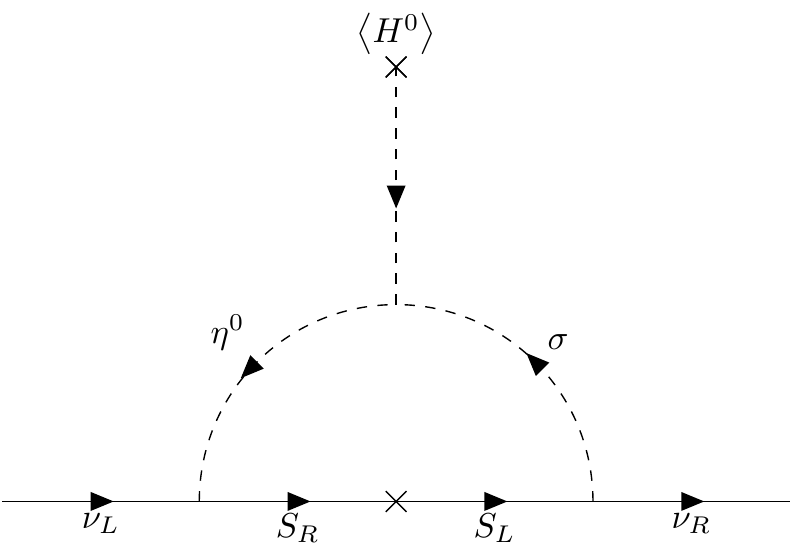}
\caption{\centering One-loop Dirac neutrino masses for Model A and B, respectively.}
\label{diag}
 \end{figure}

Due to the conservation of $B-L$ and $M_P$ neutrino masses are not generated at the tree level, arising only as a calculable one-loop contribution via the diagrams in Fig. \ref{diag}.
In both cases the neutrino masses have the same form
\be \label{nuM}
(m_\nu)_{ij} = {\sin (2\theta) \over 32 \pi^2} \sum_k y^\nu_{i k} h_{k j} m_{S_k}
\left[ {m^2_{\varphi_1} \over m^2_{\varphi_1} -
m^2_{S_k}} \ln {m^2_{\varphi_1} \over m^2_{S_k}} -
{m^2_{\varphi_2} \over m^2_{\varphi_2} -
m^2_{S_k}} \ln {m^2_{\varphi_2} \over m^2_{S_k}} \right]\,,
\ee
where, for model A, $m_{S_k}$ are the eigenvalues of the Majorana mass matrix $M_{S}^M$, while, for model B, $m_{S_k}$ are the Dirac mass matrix $M_{S}^D$ eigenvalues.
In the limit of a small $\mathbb{Z}_{2}$ soft-breaking term, {\it i.e.} $\mu_3\ll1$, we have that $\theta\ll1$, then $\sin (2\theta)\simeq \epsilon\,,$ with $\epsilon=\epsilon(\mu_3)\ll1$, 
as defined in Eq. (\ref{CNSc}).

The internal fields in the loop are odd under matter parity, while the others are even. 
The lightest among the $M_P$-odd fields is stable and, if electrically neutral, can play the role of dark matter.
Assuming that the charged component of the scalar doublet $\eta$ is heavier than the other $M_P$-odd  fields, the model can have either a complex neutral scalar or a fermion as dark matter.
In the latter case it can either be a Majorana or a Dirac fermion. 

\section{Comments on phenomenology}
\label{sec:conclusions}

The class of models suggested here has a broad range of phenomenological implications. 
Some of these are present in the minimal Stueckelberg $B-L$ extension of the \sm studied in Ref.~\cite{Heeck:2014zfa}.
For example, the existence of a $Z'$ associated to the conserved $B-L$ symmetry implies new gauge couplings of the \sm fermions, as seen in Eq. (\ref{ZpInt}). 
Hence its effects should be manifest at high energies, say, in electron-positron or proton-proton collisions. 
This implies that the ratio of the $Z'$ mass and and its corresponding gauge coupling is constrained by collider data. 
There are limits coming from Tevatron~\cite{Carena:2004xs}, LEPII and LHC~\cite{Heeck:2014zfa}. 
The most stringent current value is
\begin{equation}
M_{Z'}/g'\geq 6.9 \text{ TeV   at   95\%  C.L.}
\end{equation}

Besides the physics of the $Z'$, our proposal harbours scotogenic dark matter~\cite{Ma:2006km}, made stable by the conservation of $B-L$.
The scenario differs from the proposal in Ref.~\cite{Reig:2018mdk} 
in that the dark matter here is elementary, and can be light. The pure dark matter phenomenology has similarities with the non-scotogenic scenario discussed in~\cite{Klasen:2016qux}.
Let us first comment on the possibility of dark matter being a scalar candidate.
In this case, the lighest among the $M_P$-odd complex scalars $\varphi_1^0$ and $\varphi_2^0$ will be stable and can play the role of dark matter. 
 Consistency with direct detection experiments requires the coupling between the complex DM candidate and the $Z$-boson to be very small. 
This can be easily achieved here if the mixing between $\varphi_1^0$ and $\varphi_2^0$ is very small, and the lightest state is $\varphi_1^0$. 
In this case, the dark matter candidate $\varphi_1^0$ is mostly the scalar singlet $\sigma$,
and couples to the $Z$-boson only through its suppressed mixing with $\eta^0$. 
Notice that since the mixing angle $\theta$ is governed by the $\mathbb{Z}_2$-soft-breaking parameter $\mu_3$, 
it can be made naturally small since its absence is associated with an enhanced symmetry, and hence protected in `t Hooft sense.

The fate of mixed complex dark matter has been analysed in Ref. \cite{Kakizaki:2016dza} in a simpler phenomenological setup with no $Z'$ boson.  
In the case where only the Higgs and the $Z$-boson portals are available, the region of the parameter space compatible with the observed relic abundance and direct detection experiments is, in general, very constrained, unless co-annihilation takes place due to $\varphi_1^0$ and $\varphi_2^0$ being almost degenerate. 
The allowed region is considerably widened in the presence of a Majorana fermion, like $S_R$ in our Model A, acting as a new channel for dark matter annihilation. 
%

On the other hand, the dark matter candidate can be one of the neutral fermions, say $S_1$, if it is the lightest $M_p$-odd particle.
For the case of Model A, the dominant process contributing to the thermal relic density of the Majorana fermion dark matter candidate $S_1$ is driven by the Yukawa couplings $y^\nu$ and $h$ in Eq. (\ref{Yuk}). 
This scenario is analogous to the original scotogenic model \cite{Ma:2006km}, where fairly large Yukawas are required to produce the correct dark matter abundance, in potential tension with experimental bounds from Lepton Flavour Violation processes like $\mu\to e \gamma$ \cite{Kubo:2006yx}. 
Nevertheless, even in this case, certain regions of the parameter space are phenomenologically viable, as investigated in Ref. \cite{Vicente:2014wga}.
For the case of Model B with a Dirac fermion dark matter, there are new processes involved in setting the relic density, mediated by the $B-L$ gauge boson, according to the interactions shown in Eq. (\ref{ZpInt}). 
Assuming the $Z'$ to be the dominating channel,  the correct relic density can be successfully reproduced around the resonance condition $M_{Z'}\approx 2M_{S_1}^D$ for any $n\neq0$ in Table \ref{tabB} \cite{Han:2020oet}.
Finally, it is worth emphasising that, although some of the dark matter scenarios are more constrained than others, all of them exhibit viable regions of the parameter space.\\[-.2cm]


In short, we have proposed a simple extension of the \sm where neutrinos get naturally small ``scotogenic'' Dirac-type masses from an unbroken gauged $B-L$ symmetry.
The associated gauge boson gets mass through the Stueckelberg mechanism.
The conservation of $B-L$ and matter parity play a key role in ensuring dark matter stability.  
Dedicated studies of the resulting phenomenology will be presented elsewhere.
\acknowledgements 
\noindent

Work supported by the Spanish grants SEV-2014-0398 and FPA2017-85216-P (AEI/FEDER, UE), PROMETEO/2018/165 (Generalitat Valenciana) and the Spanish Red Consolider MultiDark FPA2017-90566-REDC. 
J. L. acknowledges financial support under grant 2019/04195-7, S\~ao Paulo Research Foundation (FAPESP). 
A.M.  Acknowledges support by CONACyT. CAV-A is supported by the Mexican C\'atedras CONACyT project 749 and SNI 58928.

\bibliographystyle{utphys}
\bibliography{bibliography}

\providecommand{\href}[2]{#2}\begingroup\raggedright\begin{thebibliography}{10}

\bibitem{Bertone:2004pz}
G.~Bertone, D.~Hooper, and J.~Silk, ``{Particle dark matter: Evidence,
  candidates and constraints},''
  \href{http://dx.doi.org/10.1016/j.physrep.2004.08.031}{{\em Phys.Rept.}
  {\bfseries 405} (2005) 279--390}.

\bibitem{Alves:2016fqe}
A.~Alves {\em et~al.}, ``{Matter-parity as a residual gauge symmetry: Probing a
  theory of cosmological dark matter},''
  \href{http://dx.doi.org/10.1016/j.physletb.2017.07.056}{{\em Phys. Lett.}
  {\bfseries B772} (2017) 825--831},
\href{http://arxiv.org/abs/1612.04383}{{\ttfamily arXiv:1612.04383 [hep-ph]}}.

\bibitem{Barbier:2004ez}
R.~Barbier {\em et~al.}, ``{R-parity violating supersymmetry},''
  \href{http://dx.doi.org/10.1016/j.physrep.2005.08.006}{{\em Phys.Rept.}
  {\bfseries 420} (2005) 1--202}.

\bibitem{Chulia:2016ngi}
S.~Centelles~Chuli\'{a} {\em et~al.}, ``{Dirac Neutrinos and Dark Matter
  Stability from Lepton Quarticity},''
  \href{http://dx.doi.org/10.1016/j.physletb.2017.01.070}{{\em Phys. Lett.}
  {\bfseries B767} (2017) 209--213},
\href{http://arxiv.org/abs/1606.04543}{{\ttfamily arXiv:1606.04543 [hep-ph]}}.

\bibitem{Chulia:2016giq}
S.~Centelles~Chuli{\'a}, R.~Srivastava, and J.~W. Valle, ``{CP violation from
  flavor symmetry in a lepton quarticity dark matter model},''
  \href{http://dx.doi.org/10.1016/j.physletb.2016.08.028}{{\em Phys.Lett.}
  {\bfseries B761} (2016) 431--436},
  \href{http://arxiv.org/abs/1606.06904}{{\ttfamily arXiv:1606.06904
  [hep-ph]}}.

\bibitem{CentellesChulia:2017koy}
S.~Centelles~Chuli{\'a}, R.~Srivastava, and J.~W.~F. Valle, ``{Generalized
  Bottom-Tau unification, neutrino oscillations and dark matter: predictions
  from a lepton quarticity flavor approach},''
  \href{http://dx.doi.org/10.1016/j.physletb.2017.07.065}{{\em Phys.Lett.}
  {\bfseries B773} (2017) 26--33},
  \href{http://arxiv.org/abs/1706.00210}{{\ttfamily arXiv:1706.00210
  [hep-ph]}}.

\bibitem{Dong:2017zxo}
P.~V. Dong {\em et~al.}, ``{The Dark Side of Flipped Trinification},''
  \href{http://dx.doi.org/10.1007/JHEP04(2018)143}{{\em JHEP} {\bfseries 04}
  (2018) 143},
\href{http://arxiv.org/abs/1710.06951}{{\ttfamily arXiv:1710.06951 [hep-ph]}}.

\bibitem{Kang:2019sab}
S.~K. Kang {\em et~al.}, ``{Scotogenic dark matter stability from gauged matter
  parity},'' \href{http://dx.doi.org/10.1016/j.physletb.2019.135013}{{\em
  Phys.Lett.} {\bfseries B798} (2019) 135013},
  \href{http://arxiv.org/abs/1902.05966}{{\ttfamily arXiv:1902.05966
  [hep-ph]}}.

\bibitem{Leite:2019grf}
J.~Leite {\em et~al.}, ``{A theory for scotogenic dark matter stabilised by
  residual gauge symmetry},''
  \href{http://dx.doi.org/10.1016/j.physletb.2020.135254}{{\em Phys. Lett.}
  {\bfseries B802} (2020) 135254}.

\bibitem{Ma:2006km}
E.~Ma, ``{Verifiable radiative seesaw mechanism of neutrino mass and dark
  matter},'' \href{http://dx.doi.org/10.1103/PhysRevD.73.077301}{{\em
  Phys.Rev.} {\bfseries D73} (2006) 077301}.

\bibitem{Farzan:2012sa}
Y.~Farzan and E.~Ma, ``{Dirac neutrino mass generation from dark matter},''
  \href{http://dx.doi.org/10.1103/PhysRevD.86.033007}{{\em Phys.Rev.}
  {\bfseries D86} (2012) 033007},
  \href{http://arxiv.org/abs/1204.4890}{{\ttfamily arXiv:1204.4890 [hep-ph]}}.

\bibitem{Hirsch:2013ola}
M.~Hirsch {\em et~al.}, ``{WIMP dark matter as radiative neutrino mass
  messenger},'' \href{http://dx.doi.org/10.1007/JHEP10(2013)149}{{\em JHEP}
  {\bfseries 1310} (2013) 149},
  \href{http://arxiv.org/abs/1307.8134}{{\ttfamily arXiv:1307.8134 [hep-ph]}}.

\bibitem{Merle:2016scw}
A.~Merle {\em et~al.}, ``{Consistency of WIMP Dark Matter as radiative neutrino
  mass messenger},'' \href{http://dx.doi.org/10.1007/JHEP07(2016)013}{{\em
  JHEP} {\bfseries 1607} (2016) 013},
  \href{http://arxiv.org/abs/1603.05685}{{\ttfamily arXiv:1603.05685
  [hep-ph]}}.

\bibitem{Hirsch:2010ru}
M.~Hirsch {\em et~al.}, ``{Discrete dark matter},''
  \href{http://dx.doi.org/10.1103/PhysRevD.82.116003}{{\em Phys.Rev.}
  {\bfseries D82} (2010) 116003},
  \href{http://arxiv.org/abs/1007.0871}{{\ttfamily arXiv:1007.0871 [hep-ph]}}.

\bibitem{Boucenna:2011tj}
M.~Boucenna {\em et~al.}, ``{Phenomenology of Dark Matter from $A_4$ Flavor
  Symmetry},'' \href{http://dx.doi.org/10.1007/JHEP05(2011)037}{{\em JHEP}
  {\bfseries 1105} (2011) 037},
  \href{http://arxiv.org/abs/1101.2874}{{\ttfamily arXiv:1101.2874 [hep-ph]}}.

\bibitem{Morisi:2012fg}
S.~Morisi and J.~W.~F. Valle, ``{Neutrino masses and mixing: a flavour symmetry
  roadmap},'' \href{http://dx.doi.org/10.1002/prop.201200125}{{\em
  Fortsch.Phys.} {\bfseries 61} (2013) 466--492},
  \href{http://arxiv.org/abs/1206.6678}{{\ttfamily arXiv:1206.6678 [hep-ph]}}.

\bibitem{Bonilla:2017ekt}
C.~Bonilla {\em et~al.}, ``{Flavour-symmetric type-II Dirac neutrino seesaw
  mechanism},'' \href{http://dx.doi.org/10.1016/j.physletb.2018.02.022}{{\em
  Phys.Lett.} {\bfseries B779} (2018) 257--261},
  \href{http://arxiv.org/abs/1710.06498}{{\ttfamily arXiv:1710.06498
  [hep-ph]}}.

\bibitem{Aranda:2013gga}
A.~Aranda {\em et~al.}, ``{Dirac neutrinos from flavor symmetry},''
  \href{http://dx.doi.org/10.1103/PhysRevD.89.033001}{{\em Phys.Rev.}
  {\bfseries D89} (2014) 033001},
  \href{http://arxiv.org/abs/1307.3553}{{\ttfamily arXiv:1307.3553 [hep-ph]}}.

\bibitem{CentellesChulia:2018gwr}
S.~Centelles~Chuli\'{a}, R.~Srivastava, and J.~W.~F. Valle, ``{Seesaw roadmap
  to neutrino mass and dark matter},''
  \href{http://dx.doi.org/10.1016/j.physletb.2018.03.046}{{\em Phys. Lett.}
  {\bfseries B781} (2018) 122--128},
\href{http://arxiv.org/abs/1802.05722}{{\ttfamily arXiv:1802.05722 [hep-ph]}}.

\bibitem{Bonilla:2018ynb}
C.~Bonilla, S.~Centelles~Chuli\'{a}, R.~Cepedello, E.~Peinado, and
  R.~Srivastava, ``{Dark matter stability and Dirac neutrinos using only
  Standard Model symmetries},''
\href{http://arxiv.org/abs/1812.01599}{{\ttfamily arXiv:1812.01599 [hep-ph]}}.

\bibitem{Ma:2019coj}
E.~Ma, ``{Leptonic Source of Dark Matter and Radiative Majorana or Dirac
  Neutrino Mass},'' \href{http://arxiv.org/abs/1912.11950}{{\ttfamily
  arXiv:1912.11950 [hep-ph]}}.

\bibitem{Reig:2018mdk}
M.~Reig {\em et~al.}, ``{Bound-state dark matter and Dirac neutrino masses},''
  \href{http://dx.doi.org/10.1103/PhysRevD.97.115032}{{\em Phys.Rev.}
  {\bfseries D97} (2018) 115032},
  \href{http://arxiv.org/abs/1803.08528}{{\ttfamily arXiv:1803.08528
  [hep-ph]}}.

\bibitem{Agostini:2019hzm}
{\bfseries GERDA} Collaboration, M.~Agostini {\em et~al.}, ``{Probing Majorana
  neutrinos with double-$\beta$ decay},''
  \href{http://dx.doi.org/10.1126/science.aav8613}{{\em Science} {\bfseries
  365} (2019) 1445}, \href{http://arxiv.org/abs/1909.02726}{{\ttfamily
  arXiv:1909.02726 [hep-ex]}}.

\bibitem{Ruegg:2003ps}
H.~Ruegg and M.~Ruiz-Altaba, ``{The Stueckelberg field},''
  \href{http://dx.doi.org/10.1142/S0217751X04019755}{{\em Int. J. Mod. Phys.}
  {\bfseries A19} (2004) 3265--3348},
\href{http://arxiv.org/abs/hep-th/0304245}{{\ttfamily arXiv:hep-th/0304245
  [hep-th]}}.

\bibitem{Pati:1974yy}
J.~C. Pati and A.~Salam, ``{Lepton Number as the Fourth Color},''
  \href{http://dx.doi.org/10.1103/PhysRevD.11.703.2}{{\em Phys.Rev.} {\bfseries
  D10} (1974) 275--289}.

\bibitem{Davidson:1978pm}
A.~Davidson, ``{$B{\ensuremath{-}}L$ as the fourth color within an
  $\mathrm{SU}(2)_L \times \mathrm{U}(1)_R \times \mathrm{U}(1)$ model},''
  \href{http://dx.doi.org/10.1103/PhysRevD.20.776}{{\em Phys.Rev.} {\bfseries
  D20} (1979) 776}.

\bibitem{Marshak:1979fm}
R.~Marshak and R.~N. Mohapatra, ``{Quark - Lepton Symmetry and B-L as the U(1)
  Generator of the Electroweak Symmetry Group},''
  \href{http://dx.doi.org/10.1016/0370-2693(80)90436-0}{{\em Phys.Lett.}
  {\bfseries B91} (1980) 222--224}.

\bibitem{Malinsky:2005bi}
M.~Malinsky, J.~Romao, and J.~W.~F. Valle, ``{Novel supersymmetric SO(10)
  seesaw mechanism},''
  \href{http://dx.doi.org/10.1103/PhysRevLett.95.161801}{{\em Phys.Rev.Lett.}
  {\bfseries 95} 161801}, \href{http://arxiv.org/abs/hep-ph/0506296}{{\ttfamily
  arXiv:hep-ph/0506296 [hep-ph]}}.

\bibitem{Feldman:2011ms}
D.~Feldman, P.~Fileviez~Perez, and P.~Nath, ``{R-parity Conservation via the
  Stueckelberg Mechanism: LHC and Dark Matter Signals},''
  \href{http://dx.doi.org/10.1007/JHEP01(2012)038}{{\em JHEP} {\bfseries 01}
  (2012) 038},
\href{http://arxiv.org/abs/1109.2901}{{\ttfamily arXiv:1109.2901 [hep-ph]}}.

\bibitem{Schechter:1980gr}
J.~Schechter and J.~W.~F. Valle, ``{Neutrino Masses in SU(2) x U(1)
  Theories},'' \href{http://dx.doi.org/10.1103/PhysRevD.22.2227}{{\em
  Phys.Rev.} {\bfseries D22} (1980) 2227}.

\bibitem{Holdom:1985ag}
B.~Holdom, ``{Two U(1)'s and Epsilon Charge Shifts},''
  \href{http://dx.doi.org/10.1016/0370-2693(86)91377-8}{{\em Phys.Lett.}
  {\bfseries B166} (1986) 196--198}.

\bibitem{Feldman:2007wj}
D.~Feldman, Z.~Liu, and P.~Nath, ``{The Stueckelberg Z-prime Extension with
  Kinetic Mixing and Milli-Charged Dark Matter From the Hidden Sector},''
  \href{http://dx.doi.org/10.1103/PhysRevD.75.115001}{{\em Phys.Rev.}
  {\bfseries D75} (2007) 115001}.

\bibitem{Williams:2011qb}
M.~Williams, C.~Burgess, A.~Maharana, and F.~Quevedo, ``{New Constraints (and
  Motivations) for Abelian Gauge Bosons in the MeV-TeV Mass Range},''
  \href{http://dx.doi.org/10.1007/JHEP08(2011)106}{{\em JHEP} {\bfseries 1108}
  (2011) 106}, \href{http://arxiv.org/abs/1103.4556}{{\ttfamily arXiv:1103.4556
  [hep-ph]}}.

\bibitem{Heeck:2014zfa}
J.~Heeck, ``{Unbroken B {\textendash} L symmetry},''
  \href{http://dx.doi.org/10.1016/j.physletb.2014.10.067}{{\em Phys.Lett.}
  {\bfseries B739} (2014) 256--262},
  \href{http://arxiv.org/abs/1408.6845}{{\ttfamily arXiv:1408.6845 [hep-ph]}}.

\bibitem{Carena:2004xs}
M.~Carena, A.~Daleo, B.~A. Dobrescu, and T.~M.~P. Tait, ``{$Z^\prime$ gauge
  bosons at the Tevatron},''
  \href{http://dx.doi.org/10.1103/PhysRevD.70.093009}{{\em Phys. Rev.}
  {\bfseries D70} (2004) 093009},
\href{http://arxiv.org/abs/hep-ph/0408098}{{\ttfamily arXiv:hep-ph/0408098
  [hep-ph]}}.

\bibitem{Klasen:2016qux}
M.~Klasen, F.~Lyonnet, and F.~S. Queiroz, ``{NLO+NLL collider bounds, Dirac
  fermion and scalar dark matter in the B{\textendash}L model},''
  \href{http://dx.doi.org/10.1140/epjc/s10052-017-4904-8}{{\em Eur.Phys.J.}
  {\bfseries C77} (2017) 348},
  \href{http://arxiv.org/abs/1607.06468}{{\ttfamily arXiv:1607.06468
  [hep-ph]}}.

\bibitem{Kakizaki:2016dza}
M.~Kakizaki, A.~Santa, and O.~Seto, ``{Phenomenological signatures of mixed
  complex scalar WIMP dark matter},''
  \href{http://dx.doi.org/10.1142/S0217751X17500385}{{\em Int. J. Mod. Phys.}
  {\bfseries A32} no.~10, (2017) 1750038},
\href{http://arxiv.org/abs/1609.06555}{{\ttfamily arXiv:1609.06555 [hep-ph]}}.

\bibitem{Kubo:2006yx}
J.~Kubo, E.~Ma, and D.~Suematsu, ``{Cold Dark Matter, Radiative Neutrino Mass,
  $\mu \to e\gamma$, and Neutrinoless Double Beta Decay},''
  \href{http://dx.doi.org/10.1016/j.physletb.2006.08.085}{{\em Phys. Lett.}
  {\bfseries B642} (2006) 18--23},
\href{http://arxiv.org/abs/hep-ph/0604114}{{\ttfamily arXiv:hep-ph/0604114
  [hep-ph]}}.

\bibitem{Vicente:2014wga}
A.~Vicente and C.~E. Yaguna, ``{Probing the scotogenic model with lepton flavor
  violating processes},'' \href{http://dx.doi.org/10.1007/JHEP02(2015)144}{{\em
  JHEP} {\bfseries 1502} (2015) 144},
  \href{http://arxiv.org/abs/1412.2545}{{\ttfamily arXiv:1412.2545 [hep-ph]}}.

\bibitem{Han:2020oet}
C.~Han, M.~L. López-Ibáñez, B.~Peng, and J.~M. Yang, ``{Dirac dark matter in
  $U(1)_{B-L}$ with Stueckelberg mechanism},''
\href{http://arxiv.org/abs/2001.04078}{{\ttfamily arXiv:2001.04078 [hep-ph]}}.

\end{thebibliography}\endgroup
\end{document}